%% file: sample-sigplan.tex
\documentclass[sigconf,twocolumn,nonacm]{acmart}
\renewcommand\footnotetextcopyrightpermission[1]{}
\AtBeginDocument{%
  }

\usepackage{multirow}
\usepackage{booktabs}
\usepackage{amsmath,amsfonts}
\usepackage{graphicx}
\usepackage{datetime}
\usepackage{tabularx}
\usepackage{makecell}

\DeclareMathAlphabet{\altmathcal}{OMS}{cmsy}{m}{n}
\newcommand*{\LastTimeAccessed}{, Accessed: \monthname[\month] \the\year}
\newcommand*{\LinkToRepo}{Link to repository: \url{https://github.com/exalsius/rca-llm}}

\begin{document}

\title{Beyond Microservices: Testing Web-Scale RCA Methods on GPU-Driven LLM Workloads}

\author{Dominik Scheinert}
\affiliation{%
  \institution{logsight.ai GmbH}
  \city{Berlin}
  \country{Germany}
}
\email{dominik.scheinert@logsight.ai}

\author{Alexander Acker}
\affiliation{%
  \institution{logsight.ai GmbH}
  \city{Berlin}
  \country{Germany}
}
\email{alexander.acker@logsight.ai}

\author{Thorsten Wittkopp}
\affiliation{%
  \institution{Technische Universität Berlin}
  \city{Berlin}
  \country{Germany}
}
\email{t.wittkopp@tu-berlin.de}

\author{Soeren Becker}
\affiliation{%
  \institution{logsight.ai GmbH}
  \city{Berlin}
  \country{Germany}
}
\email{soeren.becker@logsight.ai}

\author{Hamza Yous}
\affiliation{%
  \institution{Technology Innovation Institute}
  \city{Abu Dhabi}
  \country{UAE}
}
\email{hamza.yous@tii.ae}

\author{Karnakar Reddy}
\affiliation{%
  \institution{Technology Innovation Institute}
  \city{Abu Dhabi}
  \country{UAE}
}
\email{karnakar.reddy@tii.ae}

\author{Ibrahim Farhat}
\affiliation{%
  \institution{Technology Innovation Institute}
  \city{Abu Dhabi}
  \country{UAE}
}
\email{brahim.farhat@tii.ae}

\author{Hakim Hacid}
\affiliation{%
  \institution{Technology Innovation Institute}
  \city{Abu Dhabi}
  \country{UAE}
}
\email{hakim.hacid@tii.ae}

\author{Odej Kao}
\affiliation{%
  \institution{Technische Universität Berlin}
  \city{Berlin}
  \country{Germany}
}
\email{odej.kao@tu-berlin.de}

\begin{abstract}
\input{sections/0_abstract}
\end{abstract}



\maketitle

\input{sections/1_introduction}
\input{sections/2_towards_rca_for_llm}

\input{sections/3_methodology_rca_llm}

\input{sections/4_architecture_rca_llm}
\input{sections/5_experiment_results}

\input{sections/6_related_work}
\input{sections/7_conclusion}

\bibliographystyle{ACM-Reference-Format}
\bibliography{bib}

\end{document}

%% file: sections/0_abstract.tex
Large language model (LLM) services have become an integral part of search, assistance, and decision-making applications. 
However, unlike traditional web or microservices, the hardware and software stack enabling LLM inference deployment is of higher complexity and far less field-tested, making it more susceptible to failures that are difficult to resolve.
Keeping outage costs and quality of service degradations in check depends on shortening mean time to repair, which in practice is gated by how quickly the fault is identified, located, and diagnosed.
Automated root cause analysis (RCA) accelerates failure localization by identifying the system component that failed and tracing how the failure propagated.
Numerous RCA methods have been developed for traditional services, using request path tracing, resource metric and log data analysis.
Yet, existing RCA methods have not been designed for LLM deployments that present distinct runtime characteristics.
In this study, we evaluate the effectiveness of RCA methods on a best-practice LLM inference deployment under controlled failure injections.
Across 24 methods—20 metric-based, two trace-based, and two multi-source—we find that multi-source approaches achieve the highest accuracy, metric-based methods show fault-type-dependent performance, and trace-based methods largely fail.
These results reveal that existing RCA tools do not generalize to LLM systems, motivating tailored analysis techniques and enhanced observability, for which we formulate guidelines.

%% file: sections/1_introduction.tex
\section{Introduction}
\label{sec:introduction}

The accelerated adoption of large language models (LLMs) as a central element of search, assistance, and decision-making applications demands that model APIs are available at all times and scale with growing usage~\cite{DBLP:journals/corr/abs-2108-07258,DBLP:conf/nips/Wei0SBIXCLZ22}. 
As LLMs are increasingly deployed in critical domains such as power grids, medicine, and defense, system downtime can have severe consequences and must be minimized~\cite{DBLP:journals/artmed/NerellaBZCSBSSSBKR24,DBLP:journals/corr/abs-2307-03718}. 
This shift towards LLMs has fundamentally changed the requirements for the system infrastructure ~\cite{DBLP:journals/corr/abs-2108-07258}. 
The increasing demand for LLM APIs has forced infrastructure teams to adapt quickly, designing and scaling systems on the fly while keeping pace with rapid advances in model architectures.
Although LLM inference infrastructure designs are beginning to converge toward common industry patterns~\cite{DBLP:journals/corr/abs-2504-03648,DBLP:conf/hpca/KokolisKHKMMDSS25}, observability within these systems is still a widely underexplored area. 
Yet, observability is foundational for maintaining system reliability, availability, and security. 
Controlling failure impact and maintaining quality of service relies on minimizing mean time to repair (MTTR), which in turn depends on how quickly failures can be identified, located, and diagnosed. 
Root cause analysis (RCA) plays a critical role in this process~\cite{beyer2016sre} by enabling failures to be localized and resolved significantly faster.

Modern microservice architectures have evolved significantly, driven by the demands of scale, agility, and complexity.
Over the past decade, numerous automated RCA - primarily designed for cloud or microservices, including causality graphs~\cite{DBLP:conf/noms/WuTEK20,DBLP:conf/icsoc/LinCZ18}, statistical correlation methods~\cite{DBLP:conf/asplos/GanZHCHPD19}, and distributed tracing~\cite{DBLP:conf/sigcomm/ShenZXSLSZWY0XL23} - have emerged to diagnose and resolve issues in these systems.
LLM-based services are typically integrated into microservice architectures, providing an API for other services or users to query models.
However, while their APIs adhere to best practices, running the inference workloads requires specialized accelerator hardware and introduces unique operational characteristics, such as millisecond-scale dynamic request batching, and shared object stores for tensor transfers~\cite{DBLP:conf/sosp/KwonLZ0ZY0ZS23,DBLP:conf/osdi/MoritzNWTLLEYPJ18}. 
LLM request processing is inherently stateful, with temporary key–value caches and scheduler queues, and individual requests traversing multiple devices under model parallelism.
These properties alter observability requirements.
Traditional request-centric traces and uniform CPU or memory metrics give way to fragmented execution graphs, opaque accelerator runtimes, and hardware counters embedded within GPU drivers~\cite{DBLP:conf/hpca/KokolisKHKMMDSS25,liu2025lookbugsllminference}.

It remains unclear whether RCA techniques originally designed for microservice architectures consisting of stateless web services hold effective under such conditions~\cite{DBLP:journals/corr/abs-2407-00125,DBLP:conf/asplos/GanZHCHPD19}.
To this end, this paper presents a systematic evaluation of state-of-the-art RCA methods applied to LLM inference workloads.
We assess representative causality-graph, statistical, and trace-based tools by deploying them on an LLM inference stack aligned with best practices, and examining their performance under controlled fault injection\footnote{\LinkToRepo}.
Although some methods produce good results, our experiments show that properties of the LLM inference stack - namely novel accelerator telemetry, the lack of one-request–one-trace mappings, and complex system component interdependencies - pose generalization challenges to the evaluated RCA methods.
Therefore, we conclude by outlining telemetry enhancements and modeling adaptations necessary for effective RCA in accelerator-driven systems.
Our contributions are:

\begin{itemize}
    \item Systematic evaluation of RCA tools on LLM inference infrastructure, using controlled fault injection on distributed and representative actor-based services.
    \item Analysis of failure modes and misdiagnoses arising from mismatches between RCA assumptions and modern, quickly evolving LLM-serving systems.
    \item Guidelines for observability design to support RCA in GPU-accelerated, LLM inference deployments.
\end{itemize}

\emph{Outline}.
The remainder of the paper is structured as follows:
\autoref{sec:towards_rca_for_llm} reviews similarities and differences between LLM inference services and traditional microservices with respect to architectural considerations as well as observability requirements.
\autoref{sec:methodology_rca_llm} formulates a methodology for effective RCA in LLM inference environments and thereby narrows our research scope.
\autoref{sec:architecture_rca_llm} presents our reference stack for LLM deployments, our observability setup, and fine-grained anomaly injection procedure for controlled experiments.
\autoref{sec:experiment_results} presents and discusses our results, culminating in the formulation of guidelines.
\autoref{sec:related_work} describes the related work, while \autoref{sec:conclusion} concludes the paper.

%% file: sections/2_towards_rca_for_llm.tex
\section{LLM Inference in Microservice Architectures}
\label{sec:towards_rca_for_llm}

To achieve improved reliability for LLM inference via RCA, it is important to understand the architectural peculiarities of LLM inference stacks as well as the telemetry data needed to enable Site Reliability Engineering (SRE) operations.

\begin{figure}
    \centering
    \includegraphics[width=\columnwidth]{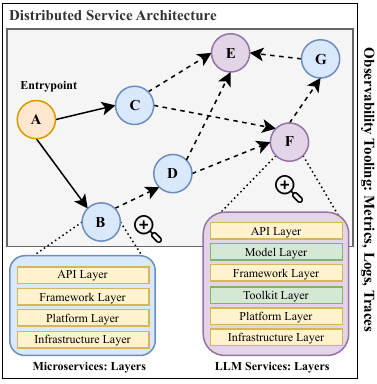}
    \caption{
    Illustration of both traditional microservices and modern LLM inference services as part of a distributed service architecture.
    We highlight similarities and differences in their individual layered architectures.
    }
    \Description{
    Illustration of both traditional microservices and modern LLM inference services as part of a distributed service architecture.
    We highlight similarities and differences in their individual layered architectures.
    }
    \label{fig:conceptual_overview}
\end{figure}

\subsection{Layered Architecture of LLM Inference Services}
\label{sec:towards_rca_for_llm_layered_architecture}

With the increased deployment and usage of LLMs or LLM-based services into traditional service architectures~\cite{moreschini2025ai,DBLP:journals/tcss/ShuZGZLLG24,DBLP:conf/ratio/LenzDSB24}, additional complexity is introduced into already heterogeneous environments composed of numerous components.
As illustrated in~\autoref{fig:conceptual_overview}, LLM inference services share similarities with traditional microservices in terms of their architecture, but also exhibit unique characteristics.
We adopt the notion of a six-layer hierarchy~\cite{DBLP:journals/corr/abs-2407-00125} and discuss similarities and differences between traditional microservices and modern LLM inference services.

At the base of the stack is the infrastructure layer, which provides the physical and virtual resources necessary for computation, including CPUs, memory, storage, networking, and specialized accelerators such as GPUs and TPUs.
While both traditional microservices and LLM services depend on this layer, microservice workloads are typically CPU-bound and rarely require specialized accelerators.
In contrast, LLM services critically depend on accelerators like GPUs and TPUs to achieve the high-throughput, low-latency inference performance needed for serving large models at scale.
The platform layer orchestrates these resources for workload execution.
In both microservice and LLM architectures, it handles scheduling, life-cycle management, and service abstraction, often realized through orchestration systems like Kubernetes.
However, LLM services extend this layer with distributed ML compute frameworks (e.g., Ray) that are specialized for scaling model execution across multiple GPUs or nodes.
While microservices typically operate as independent units with minimal cross-node coordination, LLM workloads require such frameworks to partition and coordinate model execution when a single node’s resources are insufficient.

The toolkit layer provides runtime libraries and operator integrations that enable applications to utilize specialized hardware.
This layer is specific to LLM services because they require direct access to GPU compute capabilities via libraries such as CUDA and cuDNN, as well as hardware operators (e.g., from NVIDIA, AMD), which install and manage these dependencies within the platform.
In contrast, traditional microservices do not require these toolkits because they execute on general-purpose CPUs without interacting with vendor-specific accelerator drivers.
The framework layer provides a structured approach for implementing the actual application.
While traditional microservices use general-purpose application frameworks to implement business logic (e.g., Spring Boot), LLM services rely on AI frameworks (e.g., Hugging Face Transformers).
These provide model-definition languages, automatic differentiation, and optimized inference kernels.
Thus, although both architectures include a framework layer, their purpose differs fundamentally: business logic execution in microservices is contrasted by efficient tensor computation and model inference in LLM stacks.

The model layer encapsulates artifacts such as parameter checkpoints, tokenizer vocabularies, and prompt templates that define the learned function of an LLM.
This layer has no direct analogue in traditional microservices, which do not manage learned models as operational artifacts.
Instead, their deployable units are usually code binaries or container images with static configurations instead of learned parameters.
Finally, the API layer exposes functionality to external consumers.
Here, both traditional microservices and LLM inference services converge architecturally, as both provide user-facing APIs, handle request admission, perform routing, and enforce access control.
However, LLM services integrate specialized inference logic (e.g., via vLLM), such as continuous batching, token streaming, cache-aware routing, and context window management, which extend beyond the oftentimes stateless request-response patterns typical of microservice APIs in modern service architectures.

Although real-world services sometimes blur these boundaries, for example, frameworks may ship their own serving gateways or toolkits may influence placement decisions, this layered view clarifies responsibilities across architectures.
In both paradigms, the infrastructure provides resources and the platform manages them, but LLM inference introduces distinct toolkits, ML frameworks, and model layers, with the API layer exposing them as a unified endpoint.

\subsection{Requirements for Observability}
\label{sec:towards_rca_for_llm_observability}

At the core of observability are three complementary signal types that enable the diagnostic of system health and recovery in case of failures.
For LLM services, we revisit the core challenges and key considerations to establish observability.

Metrics capture continuous, machine-readable measurements of system behavior, such as CPU utilization, memory consumption, network throughput, and request latency.
In traditional microservices, metrics primarily track resource usage, request rates, and error counts at the service or container level, enabling trend analysis and threshold-based alerting with minimal storage cost.
With modern LLM inference services, additional metrics are exposed or qualify for analysis, including GPU-centric metrics (e.g., memory bandwidth usage, kernel execution time) and specialized indicators such as time-to-first-token (TTFT) and tokens generated per second.
These metrics are essential to identify performance bottlenecks in accelerator-based inference workloads, which are not observable using CPU-only telemetry.

Logs record discrete events, warnings, and error messages, providing narrative context beyond the numerical aggregates of metrics.
In microservices, logs typically include HTTP request logs, database queries, and stack traces, which can be correlated via request IDs or trace contexts to reconstruct failure scenarios.
As for LLM inference services, they typically produce logs with different semantics, including model loading events, GPU allocation decisions, tokenization errors, and framework-level warnings (e.g., CUDA kernel failures).
Ensuring logs are enriched with relevant metadata, such as model identifiers, prompt lengths, and GPU device IDs, is critical to diagnose failures unique to ML-serving workloads.

Distributed traces stitch these events and metrics into an end-to-end representation of request execution, capturing the causal path through services, gateways, and processing stages.
In microservices, traces map multi-hop service dependencies, database queries, and inter-service calls, supporting diagnosis of latency spikes and bottlenecks along the request path.
For LLM inference services, traces often remain confined within the inference service boundary, capturing intra-node execution flows such as request queuing, batching decisions, model invocation, and GPU execution spans.
This is because the entire inference workflow either runs inside a single service with limited cross-node-service calls, or its distributed execution is concealed by recent technologies like NVLink, challenging traditional tracing methodologies.
Thus, LLM traces must expose fine-grained intra-service activity rather than inter-service dependencies.

Collecting these telemetry signals in both architectures requires minimal-intrusion instrumentation to avoid performance degradation.
Once emitted, metrics, logs, and traces converge in observability platforms to enable downstream data analysis tasks.
For instance, when an anomaly such as elevated TTFT is detected during LLM inference, engineers typically begin with the observed metric deviation, then correlate it with GPU resource utilization metrics, review framework and driver logs for potential errors, and analyze trace spans to pinpoint contributing factors such as queuing delays or suboptimal batching behavior.
Overall, while the conceptual pillars of observability remain consistent between microservices and LLM inference services, the nature, granularity, and interpretation of these signals may differ substantially.
Microservices primarily focus on inter-service latency and availability, whereas LLM workloads demand telemetry that also captures accelerator utilization, framework-level execution, and model-specific performance indicators.
With regard to~\autoref{fig:conceptual_overview}, comprehensive observability for LLM inference services should integrate signals across all layers, from infrastructure and platform to toolkits, frameworks, models, and APIs, to enable a holistic understanding and targeted mitigation of occurring faults.

%% file: sections/3_methodology_rca_llm.tex
\section{RCA Methodology for LLM Inference}
\label{sec:methodology_rca_llm}

Production-grade LLM inference systems are distributed systems by design.
Ensuring distributed system dependability requires efficiently detecting, localizing, and mitigating failures.
Dependability encompasses attributes such as availability, reliability, and maintainability~\cite{DBLP:journals/pieee/AvizienisL86}, with availability commonly defined as the fraction of time a system delivers correct service:

\begin{equation*}
    \text{Availability} = \frac{\text{MTBF}}{\text{MTBF} + \text{MTTR}}.
\end{equation*}

Here, MTBF denotes mean time between failures and MTTR denotes mean time to repair. 
Thus, reducing MTTR  becomes an operational objective~\cite{DBLP:series/synthesis/2009Barroso} for production systems.

Failures in distributed systems rarely stay confined to a single component.
Due to interdependencies, localized faults may propagate, resulting in cascading failures that degrade system-wide performance or availability~\cite{DBLP:journals/jss/CotroneoPRT16}.
Rapid identification of the origin of such failures is essential to limit their impact.
Site reliability engineering (SRE) operationalizes this objective through practices that leverage observability data to detect, localize, and mitigate failures rapidly~\cite{beyer2016sre}.

Within this workflow, root cause analysis (RCA) plays an important role. 
RCA seeks to identify the component or group of components whose malfunction initiated an observed failure (e.g., slow token generation).
Formally, let a distributed system consist of \(N\) components 
\(\mathcal{C}= \{c_1,\dots,c_N\}\).
Each component \(c_i\) emits \(M\) telemetry signals over time $t$, yielding a
multivariate time series $\mathbf{X}^i_t \;=\; \bigl(x^{(i,1)}_t,\;\dots,\;x^{(i,M)}_t\bigr).$
Given an observed failure at time $t_F$, RCA operates on the telemetry data within an observation window $[t_0, t_\mathrm{rca}]$ with \(t_0 < t_F < t_{\mathrm{rca}}\), aiming to infer the subset of components $c^* \subseteq {C}$ that caused the failure.
In practice, $t_0$ is chosen to keep telemetry data manageable for RCA methods or by implicit dictate of time series database retention policies.

Many automated RCA approaches construct an interdependency graph $\mathcal{G} = (V,E)$ over metrics, logs, or traces to model system structure and information flow~\cite{DBLP:conf/kbse/PhamH024}. 
A scoring function $s: V \to \mathbb{R}$ is then applied to prioritize nodes (e.g., metrics or components) most likely to explain the failure event. 
Generally, RCA is challenged by high-dimensional, sparsely labeled data, limited supervision, temporal delays between cause and effect, and incomplete observability.
Effective RCA techniques must therefore be able to infer causal structure or correlations from noisy, partial telemetry.

\subsection{Research Objective}
\label{sec:methodology_rca_llm_research_objective}

While RCA has been extensively explored for traditional microservice applications, its applicability to modern LLM inference services remains largely unexamined.
As previously outlined, these services can differ along various dimensions.
Yet, existing RCA methods have been primarily designed with traditional microservices in mind.
It is unclear whether the assumptions and methodologies of existing RCA methods hold in LLM-serving environments, where failures may surface in other ways due to architectural differences.
This motivates our study to systematically evaluate 24 RCA methods, testing their performance to identify the correct root cause for four fault scenarios within an LLM inference stack that conforms to best practices in industry and research.

%% file: sections/4_architecture_rca_llm.tex
\section{Enabling RCA for LLM Inference Architectures}
\label{sec:architecture_rca_llm}

\begin{figure}
    \centering
    \includegraphics[width=0.8\columnwidth]{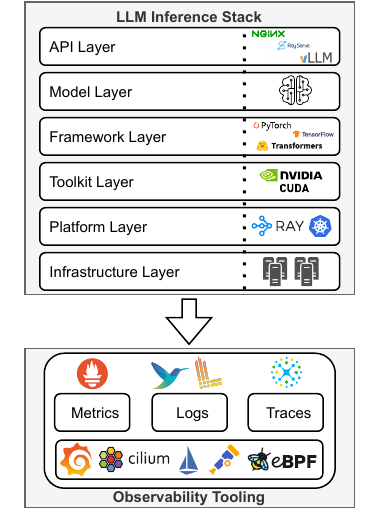}
    \caption{
    High-level overview of our LLM inference deployment stack and observability tooling.
    The different architecture layers as well as telemetry data pillars, previously discussed in~\autoref{sec:towards_rca_for_llm}, are mapped to concrete technologies, eventually forming our experimental system and serving as a blueprint for other practitioners and researchers.
    }
    \Description{
    High-level overview of our LLM inference deployment stack and observability tooling.
    The different architecture layers as well as telemetry data pillars, as discussed in the previous section, are mapped to concrete technologies, eventually forming our experimental system and serving as a blueprint for other practitioners and researchers.
    }
    \label{fig:reference_stack}
\end{figure}

As LLM serving continues to be an emerging field with few standardized solutions, we developed an experimental system\footnote{\LinkToRepo} that implements current best practices~\cite{DBLP:conf/osdi/MoritzNWTLLEYPJ18,DBLP:journals/corr/abs-2504-03648,DBLP:conf/sosp/KwonLZ0ZY0ZS23} for scalable inference while supporting detailed observability and fault injection.
This system provides a realistic foundation for evaluating RCA methods under controlled conditions.
In the following, we describe its architecture, mapping it to the LLM inference layers previously discussed, and detail the integrated observability tooling, and automated anomaly injection workflow used to enable systematic assessment.
Where feasible, we use Ansible routines and community-maintained Helm charts for reproducibility.
A general high-level overview is provided with~\autoref{fig:reference_stack}.

\subsection{LLM Deployment Reference Stack}
\label{sec:architecture_rca_llm_reference_stack}

For the different layers of a typical LLM inference architecture, we choose the following technologies:

\textbf{Infrastructure.}
Our infrastructure is built on eight leased virtual machines (VMs) from a public cloud provider, each equipped with 32 vCPUs, one NVIDIA A100 PCIe 80GB GPU, and 240GB RAM, which is a representative configuration for modest self-hosted LLM clusters: Recent peer-reviewed systems studies evaluate LLM serving on single-A100 80GB hosts and small clusters~\cite{DBLP:conf/mlsys/Chen0WZCK24,DBLP:conf/mlsys/0007CLHLYCZZK0S24}, and use A100 80GB cloud instances in production-like experiments~\cite{DBLP:conf/osdi/AgrawalKPMKGTR24,DBLP:conf/osdi/YuJKKC22}.
This allows us to experiment with dynamic resource allocation while supporting data and model parallelism.

\textbf{Platform.}
All machines are orchestrated using Kubernetes~\cite{DBLP:conf/eurosys/VermaPKOTW15}, specifically the lightweight K3s\footnote{\url{https://k3s.io/}\LastTimeAccessed} distribution.
Kubernetes manages compute (Pods, Deployments), storage (Volumes, Secrets, ConfigMaps), and networking (Services, Ingress) resources.
Moreover, the platform layer is implemented by integrating Kubernetes with Ray~\cite{DBLP:conf/osdi/MoritzNWTLLEYPJ18}, an open-source distributed compute framework that provides scalable APIs for data processing, training, and inference.
Ray integrates tightly with Kubernetes via the kuberay-operator\footnote{\url{https://github.com/ray-project/kuberay}\LastTimeAccessed}, which defines custom Kubernetes resources such as RayCluster and RayService.
A RayCluster defines the composition and scaling of Ray Pods (head and workers), while RayService couples cluster configuration with deployment logic.
We expose services created by a RayService externally via an NGINX gateway.
This configuration supports distributed LLM inference across multi-node, multi-GPU environments.

\textbf{Toolkit.}
We use the NVIDIA GPU Operator\footnote{\url{https://github.com/NVIDIA/gpu-operator}\LastTimeAccessed} to automate the provisioning of CUDA, drivers, and runtime components required for GPU support.
The operator ensures correct discovery and configuration of GPUs within Kubernetes, simplifying this typically error-prone process.

\textbf{Framework.}
We rely on PyTorch-based tooling and, in particular, Hugging Face Transformers to interface with pre-trained models.
Although not directly selected in our reference stack, Transformers~\cite{DBLP:conf/emnlp/WolfDSCDMCRLFDS20} is the underlying library used by our inference engine vLLM for handling models.

\textbf{Model.}
Model management~\cite{DBLP:journals/corr/abs-2407-00125} is handled through Hugging Face repositories, which provide access to pre-trained LLMs from providers like TII Falcon, Qwen or Llama.

\textbf{API.}
The API layer is built using vLLM components, a fast and efficient serving library for Transformer-based models.
vLLM~\cite{DBLP:conf/sosp/KwonLZ0ZY0ZS23} supports advanced features such as continuous batching and OpenAI-compatible APIs.
It integrates with Ray Serve to enable distributed serving and inference.
The NGINX gateway routes incoming requests to Ray Serve Pods based on the model ID in the URL path.
Requests are distributed across available replicas, each running a vLLM engine instance responsible for pre-processing, inference, and post-processing.
This setup provides scalable, multi-node LLM serving with high throughput.

\subsection{Observability for LLM Inference Deployments}
\label{sec:architecture_rca_llm_observability}

We extend our LLM inference stack with an observability solution covering metrics, logs, and traces, and integrate it with our Kubernetes-based deployment.
We also use Kubernetes with the eBPF-based Cilium\footnote{\url{https://cilium.io/}\LastTimeAccessed} Container Network Interface (CNI) plugin for enhanced network-level observability.

\textbf{Metrics.}
We collect metrics using Prometheus\footnote{\url{https://prometheus.io/}\LastTimeAccessed} and associated tools.
This setup captures both system-level metrics (e.g., CPU, memory) and application-specific metrics from Ray and vLLM.
Prometheus scrapes exposed endpoints defined by Kubernetes PodMonitor and ServiceMonitor resources.
We also deploy Istio\footnote{\url{https://istio.io/}\LastTimeAccessed} as a service mesh using its ambient deployment mode (ztunnel + waypoint proxy) and collect its telemetry to monitor network communication between LLM deployment components.
This includes metrics on request rates, latencies, and errors across services.
Collected key LLM inference signals such as TTFT and E2E latency are visualized using Grafana\footnote{\url{https://grafana.com/}\LastTimeAccessed}.
Metrics are enriched with Kubernetes metadata for fine-grained filtering and debugging.

\textbf{Logs.}
Application logs from Ray and the NGINX gateway are captured via Fluent Bit\footnote{\url{https://fluentbit.io/}\LastTimeAccessed} sidecar containers.
Sidecars read logs from shared volumes and forward them to Loki\footnote{\url{https://grafana.com/oss/loki/}\LastTimeAccessed} for persistence.
This approach avoids any modification to application code and supports structured log aggregation.
Logs are tagged with metadata (e.g., Pod name, namespace), enabling effective correlation with metrics and traces.

\textbf{Traces.}
To capture distributed traces without instrumenting application code, we use DeepFlow~\cite{DBLP:conf/sigcomm/ShenZXSLSZWY0XL23}, an eBPF-based tracing solution.
DeepFlow agents run as DaemonSets in the Kubernetes cluster and attach to kernel-level events.
This enables zero-intrusion tracing of end-to-end request paths with minimal overhead~\cite{9527003}.
Trace data is enriched with metadata (e.g., HTTP headers like X-Request-ID) and visualized in Grafana via custom plugins.
This allows us to localize failures across components in distributed LLM serving pipelines.

\begin{figure}
    \centering
    \includegraphics{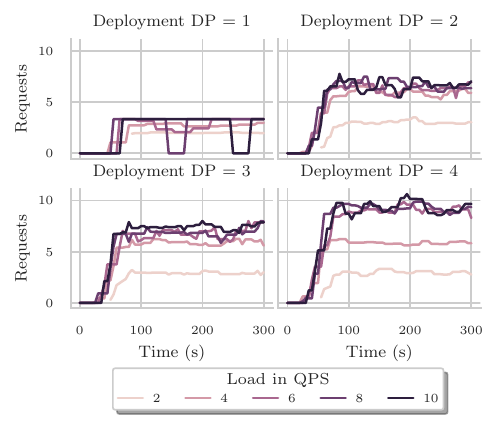}
    \caption{
    We present exemplary results of conducted smoke tests that show the number of processed requests over time when facing a constant load of queries per second (QPS) in light of different data parallelism (DP) configurations.
    }
    \Description{
    Validating our reference stack and observability tooling: We exemplary show the number of successfully processed requests over time when facing a constant load.
    }
    \label{fig:smoke_test_load_and_metrics}
\end{figure}

Overall, our cloud-native observability stack supports experimentation and troubleshooting in production-like LLM deployments.
Furthermore, it enables the collection of telemetry data that fuels our systematic RCA method assessment.
As validated by a series of smoke tests, one of which its results are displayed in~\autoref{fig:smoke_test_load_and_metrics}, the compiled LLM inference stack is capable of handling non-trivial load, all while being comprehensively analyzed by our observability solution.

\subsection{Load Generation and Failure Injections}
\label{sec:architecture_rca_llm_anomaly_injection}

For our systematic evaluation of RCA tools on LLM inference infrastructure, we need to run realistic load scenarios while injecting controlled faults into deployed inference services.

\textbf{Target System.}
Without loss of generality, we deploy the Falcon-H1-7B-Instruct LLM within our inference stack described in~\autoref{sec:architecture_rca_llm_reference_stack}, where each Pod of the RayCluster has access to one GPU, eight CPU cores, and 16GB of memory.
In total, the RayCluster is composed of one Ray head Pod and three Ray worker Pods.
The model is deployed with a data parallelism of four, which means that each Pod of the RayCluster manages and runs a model replica.
Notably, the concrete LLM selection is irrelevant for the scope of our research, as we investigate the LLM inference stack as a whole and treat served models as interchangeable workloads.

\begin{figure}
    \centering
    \includegraphics[width=\columnwidth]{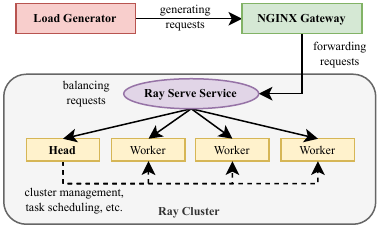}
    \caption{
    Abstract illustration of the setup and request handling workflow.
    A load generator sends request to the NGINX gateway that is exposed externally from the cluster.
    The gateway then forwards request to the Kubernetes service responsible for balancing serve requests for a particular LLM.
    Eventually, this service sends the request to one of the available endpoints, which includes a single Ray head Pod and any number of connected Ray worker Pods.
    }
    \Description{
    Abstract illustration of the setup and request handling workflow.
    A load generator sends request to the NGINX gateway that is exposed externally from the cluster.
    The gateway then forwards request to the Kubernetes service responsible for balancing serve requests for a particular LLM.
    Eventually, this service sends the request to one of the available endpoints, which includes a single Ray head Pod and any number of connected Ray worker Pods.
    }
    \label{fig:experiment_setup}
\end{figure}

\textbf{Load Generation.}
To evaluate the system under load, we use a custom benchmark tool built on Locust\footnote{\url{https://locust.io/}\LastTimeAccessed}, targeting a constant query rate of five Queries Per Second (QPS) using up to 300 workers.
Each test runs for 360 seconds and generates a stream of chat-based requests to the inference endpoint of the deployed LLM.
Each query contains on average 512 prompt tokens and results in 256 streamed output tokens, representing a typical Q\&A chat~\cite{DBLP:conf/icml/ChiangZ0ALLZ0JG24}, matching the OpenAI-compatible API used in our deployment.
To limit request concurrency, we cap the number of simultaneously running requests at 500.
This setup, illustrated in~\autoref{fig:experiment_setup}, allows us to establish a performance baseline with regard to system scalability, throughput, and latency under controlled load conditions, before we continue to study how the observability stack and deployed RCA methods respond to synthetically created performance bottlenecks.

\textbf{Anomaly Injection.}
We implement three Pod-level fault injection experiments from related work~\cite{DBLP:conf/noms/WuTEK20} and integrate them into our LLM inference stack using ChaosMesh\footnote{\url{https://chaos-mesh.org/}\LastTimeAccessed}.
Additionally, we implement a GPU-level fault inducing GPU stress, resulting in the following four fault scenarios:
\begin{itemize}
    \item \textbf{CPU Hog}: We stress with 16 threads, spawned in the target Pod, at full load (StressChaos CRD).
    \item \textbf{Memory Leak}: We allocate 8GB of additional memory over time (StressChaos CRD) in the target Pod.
    \item \textbf{Network Latency}: Delay all incoming traffic to the target Pod by one second (NetworkChaos CRD).
    \item \textbf{GPU Throttling}: Throttling the GPU attached to the target Pod by setting the power limit from 300W to 150W using the nvidia-smi utility tool.
\end{itemize}
We inject faults sequentially into the Ray worker Pods, each over a period of one minute, with cooldown periods in-between to restore normal behavior.
While failures in real-world deployments may manifest concurrently (e.g., GPU throttling coinciding with network latency), we follow established dependability benchmarking practice by injecting one fault at a time.
Sequential single-fault injection ensures experimental control, reproducibility, and clear attribution of observed anomalies to injected causes.
Prior work in fault injection and chaos engineering similarly emphasizes single-fault experiments as the methodological baseline before considering more complex, multi-fault scenarios~\cite{DBLP:journals/tdsc/AvizienisLRL04,DBLP:journals/computer/HsuehTI97,DBLP:journals/software/BasiriBRHKRR16}.
This setup enables us to create a rich dataset using our previously discussed observability system.
Once the dataset is captured, our goal is to study how established RCA methods generalize to the unique telemetry and failure patterns of LLM systems.

%% file: sections/5_experiment_results.tex
\section{Experiment Results}
\label{sec:experiment_results}

In this section, we present the results of our experiments\footnote{\LinkToRepo}, analyzing the performance and limitations of existing RCA methods in light of selected evaluation metrics when applied to LLM inference workloads under controlled fault scenarios.

\subsection{Evaluation Metrics}
\label{sec:evaluation_metrics}

To quantify the performance of the selected baselines on a set of anomalies $\mathcal{A}$ injected into an LLM inference system, we use two standard metrics, $AC@k$ and $Avg@k$, from the literature.
The former metric denotes the probability that the top-$k$ results returned by an algorithm include the actual root cause.
A higher $AC@k$ score, particularly for small values of $k$, indicates that the algorithm correctly identifies the root cause early in the ranking.
For an anomaly $a\in\mathcal{A}$, let $R^a[i]$ denote the $i$-th ranked service and $v^a_{\mathrm{rc}}$ be the set of true root causes.
Then, $AC@k$ is defined as:
\begin{equation}
    AC@k = \frac{1}{|\mathcal{A}|} \sum_{a \in \mathcal{A}} \frac{\sum_{i < k} \left(R^a[i] \in v^a_{\mathrm{rc}}\right)}{\min(k, |v^a_{\mathrm{rc}}|)}
\end{equation}
Consequently, $Avg@k$ measures the overall performance of an RCA method and is defined as $Avg@k = \frac{1}{k}\sum_{j=1}^{k} AC@j$.
For the interpretation of these evaluation metrics, note that RCA methods rank among six candidate Pods (three Ray workers, one Ray head, one NGINX gateway, and one KubeRay operator), yielding a random baseline of $AC@1 \simeq \frac{1}{6}$.

\subsection{Performance Analysis of RCA Methods}
\label{sec:experiment_results_description}

\begin{table*}
\centering
\caption{RCA performance of 24 different baselines across four fault types.}
\resizebox{\textwidth}{!}{
\begin{tabular}{c|cccc|ccc|ccc|ccc}
\toprule
\multirow{2}{*}{\makecell{\textbf{Data} \\ \textbf{Source}}} & \multirow{2}{*}{\textbf{Method}} 
& \multicolumn{3}{c|}{\textbf{CPU}} 
& \multicolumn{3}{c|}{\textbf{MEMORY}} 
& \multicolumn{3}{c}{\textbf{NETWORK}}
& \multicolumn{3}{c}{\textbf{GPU}} \\
\cmidrule(lr){3-5} \cmidrule(lr){6-8} \cmidrule(lr){9-11} \cmidrule(lr){12-14}
& & AC@1 & AC@3 & Avg@5 & AC@1 & AC@3 & Avg@5 & AC@1 & AC@3 & Avg@5 & AC@1 & AC@3 & Avg@5 \\
\midrule
\multirow{20}{*}{Metric}
& Random           & 0.17 & 0.50 & 0.51 & 0.10 & 0.43 & 0.47 & 0.30 & 0.57 & 0.58 & 0.23 & 0.63 & 0.57 \\
& PC-PR            & 0.10 & 0.43 & 0.38 & 0.20 & 0.53 & 0.52 & 0.07 & 0.47 & 0.45 & 0.33 & \textbf{0.67} & 0.60 \\
& PC-RW            & 0.33 & 0.57 & 0.66 & 0.33 & 0.47 & 0.63 & 0.33 & 0.60 & 0.66 & 0.33 & 0.43 & 0.61 \\
& FCI-PR           & 0.23 & 0.53 & 0.57 & 0.17 & 0.40 & 0.44 & 0.33 & 0.53 & 0.59 & 0.07 & 0.53 & 0.49 \\
& FCI-RW           & 0.33 & 0.57 & 0.66 & 0.33 & 0.47 & 0.63 & 0.33 & 0.60 & 0.66 & 0.33 & 0.43 & 0.61 \\
& Granger-PR       & 0.23 & 0.37 & 0.45 & 0.13 & 0.57 & 0.53 & 0.20 & 0.57 & 0.57 & 0.23 & 0.57 & 0.55 \\
& Granger-RW       & 0.33 & 0.57 & 0.66 & 0.33 & 0.47 & 0.63 & 0.33 & 0.60 & 0.66 & \textbf{0.37} & 0.53 & \textbf{0.65} \\
& LiNGAM-PR     & 0.00 & 0.03 & 0.09 & 0.00 & 0.13 & 0.23 & 0.00 & 0.17 & 0.25 & 0.10 & 0.53 & 0.49 \\
& LiNGAM-RW     & 0.33 & 0.57 & 0.66 & 0.33 & 0.47 & 0.63 & 0.33 & 0.60 & 0.66 & 0.33 & 0.43 & 0.61 \\
& CausalRCA~\cite{DBLP:journals/jss/XinCZ23} & 0.20 & 0.57 & 0.59 & 0.20 & 0.40 & 0.43 & 0.17 & 0.40 & 0.45 & 0.23 & 0.57 & 0.57 \\
& CausalAI~\cite{DBLP:journals/corr/abs-2301-10859} & 0.33 & 0.47 & 0.43 & 0.37 & 0.50 & 0.48 & 0.23 & 0.23 & 0.23 & 0.03 & 0.07 & 0.05 \\
& MicroCause~\cite{DBLP:conf/iwqos/MengZSZHZJWP20} & 0.00 & 0.07 & 0.12 & 0.00 & 0.07 & 0.14 & 0.00 & 0.07 & 0.13 & 0.03 & 0.20 & 0.27 \\
& $\epsilon$-Diagnosis~\cite{DBLP:conf/www/ShanCLZXHLD19} & 0.27 & 0.70 & 0.57 & 0.20 & 0.43 & 0.37 & 0.40 & 0.70 & 0.63 & 0.17 & 0.53 & 0.45 \\
& BARO~\cite{DBLP:journals/pacmse/PhamH024} & 0.13 & \textbf{1.00} & 0.83 & 0.77 & 0.93 & \textbf{0.91} & 0.33 & \textbf{1.00} & 0.87 & 0.20 & \textbf{0.67} & 0.59 \\
& RCD~\cite{DBLP:conf/nips/IkramCMSBK22} & 0.50 & 0.80 & 0.73 & 0.43 & 0.77 & 0.69 & 0.43 & 0.83 & 0.73 & 0.23 & 0.53 & 0.46 \\
& CIRCA~\cite{DBLP:conf/kdd/0005LYNZSP22} & 0.40 & \textbf{1.00} & 0.86 & \textbf{0.80} & 0.93 & \textbf{0.91} & 0.70 & \textbf{1.00} & 0.93 & 0.20 & 0.63 & 0.61 \\
& NSigma~\cite{DBLP:conf/icsoc/LinCZ18} & \textbf{1.00} & \textbf{1.00} & \textbf{1.00} & 0.70 & 0.90 & 0.87 & \textbf{0.93} & \textbf{1.00} & \textbf{0.99} & 0.17 & 0.63 & 0.55 \\
& MicroRCA~\cite{DBLP:conf/noms/WuTEK20} & 0.20 & 0.50 & 0.51 & 0.13 & 0.37 & 0.46 & 0.60 & 0.90 & 0.81 & 0.07 & 0.50 & 0.42 \\
& Microscope~\cite{DBLP:conf/icsoc/LinCZ18} & 0.20 & 0.73 & 0.56 & 0.30 & 0.87 & 0.65 & 0.63 & 0.90 & 0.82 & 0.13 & 0.53 & 0.41 \\
& MonitorRank~\cite{DBLP:conf/sigmetrics/KimSS13} & 0.13 & 0.27 & 0.33 & 0.03 & 0.10 & 0.21 & 0.27 & 0.77 & 0.69 & 0.07 & 0.43 & 0.35 \\
\midrule
\multirow{2}{*}{Trace}
& MicroRank~\cite{DBLP:conf/www/YuCCGHJWSL21} & 0.00 & 0.00 & 0.20 & 0.00 & 0.00 & 0.20 & 0.00 & 0.00 & 0.20 & 0.00 & 0.00 & 0.20 \\
& TraceRCA~\cite{DBLP:conf/iwqos/Li0JZWZWJYWCZNS21} & 0.03 & 0.67 & 0.55 & 0.00 & 0.00 & 0.01 & 0.00 & 0.30 & 0.28 & 0.00 & 0.20 & 0.15 \\
\midrule
\multirow{2}{*}{\makecell{Multi \\ Source}}
& MM-BARO~\cite{DBLP:journals/pacmse/PhamH024} & 0.13 & \textbf{1.00} & 0.83 & 0.77 & 0.93 & 0.89 & 0.33 & \textbf{1.00} & 0.87 & 0.10 & 0.37 & 0.41 \\
& PDiagnose~\cite{DBLP:conf/ispa/HouJWLH21} & 0.00 & \textbf{1.00} & 0.79 & 0.00 & \textbf{1.00} & 0.80 & 0.00 & 0.87 & 0.71 & 0.00 & 0.57 & 0.58 \\
\bottomrule
\end{tabular}
}
\label{tbl:experiment_results}
\end{table*}

We build upon an open-source RCA benchmark that has also been proposed and used in recent publications~\cite{DBLP:conf/kbse/PhamH024,DBLP:conf/www/Pham00S025}.
It features a variety of state-of-the-art RCA methods -- most of them operating on metrics (e.g., CIRCA~\cite{DBLP:conf/kdd/0005LYNZSP22}, CausalAI~\cite{DBLP:journals/corr/abs-2301-10859}), but some also on traces (e.g., TraceRCA~\cite{DBLP:conf/iwqos/Li0JZWZWJYWCZNS21}, MicroRank~\cite{DBLP:conf/www/YuCCGHJWSL21}) or even conducting a multi-source approach (e.g., multi-source BARO~\cite{DBLP:journals/pacmse/PhamH024}, PDiagnose~\cite{DBLP:conf/kdd/0005LYNZSP22}).
We adapt this benchmark framework to also consume and handle our own datasets derived from collected telemetry data during our experiments.
In addition, we integrate three more RCA methods: MicroRCA~\cite{DBLP:conf/noms/WuTEK20}, Microscope~\cite{DBLP:conf/icsoc/LinCZ18}, and MonitorRank~\cite{DBLP:conf/sigmetrics/KimSS13}, all of which are also commonly used in the related work for comparison.
This gives us a total of 24 baselines.
Each RCA method receives \raisebox{.5pt}{\textcircled{\raisebox{-.9pt}{1}}} telemetry from the five-minute window preceding the injected anomaly, \raisebox{.5pt}{\textcircled{\raisebox{-.9pt}{2}}} the injection timestamp, and \raisebox{.5pt}{\textcircled{\raisebox{-.9pt}{3}}} telemetry from the one-minute injection interval.

We present the main results of our experiments in~\autoref{tbl:experiment_results}.
Here, we summarize the results of all baselines across the different fault types and in terms of previously introduced evaluation metrics.
Each fault type has been injected ten times into each of the three Ray worker Pods, yielding a total of 30 injections per fault type and a total of 120 injections across fault types.
Note that, apart from a random selection strategy and all the explicitly named RCA methods, we also include a variety of methods that are effectively a combination of causal discovery methods (Peter-Clark (PC)~\cite{DBLP:books/daglib/0023012}; Fast Causal Inference (FCI)~\cite{DBLP:books/daglib/0023012}; Granger~\cite{Granger1980TestingFC}; LiNGAM~\cite{DBLP:journals/jmlr/ShimizuHHK06}) and scoring methods (PageRank (PR); Random Walk (RW)).
In the rare case of an RCA method failing to converge, we handle this error by defaulting to the random selection strategy.

\paragraph{Metric-based Methods}
In line with the utilized RCA benchmark and related work, metric-based methods are supplied general metrics on CPU utilization, memory consumption, network traffic, and latency per Pod.
In addition, to account for the peculiarities of AI workloads, we include information on GPU utilization.
In case the respective RCA method supports it, we also provide a Directed Acyclic Graph (DAG) object derived from network communication to convey a broad understanding of dependencies between Pods.

Table~\ref{tbl:experiment_results} shows that metric-based RCA methods yield mixed performance across fault types. 
Classical causal discovery approaches such as PC-PR, FCI-PR, and Granger-PR achieve moderate $Avg@5$ scores (e.g., FCI-PR: $0.57$ CPU, $0.59$ network), but their $AC@1$ remains low ($\leq 0.33$), indicating that while the true root cause is often within the top-5 candidates (out of 6 total Pods), it rarely ranks first. 
This suggests these methods may lack decisive ranking capability, an observation consistent with prior evaluations in microservice systems.
In contrast, modern methods such as NSigma achieve near-perfect performance, with $AC@1=1.00$ for CPU and $\geq 0.70$ for memory and network. 
Similarly, BARO, CIRCA, and RCD exhibit strong ranking accuracy, e.g., CIRCA reaches $AC@1$ of 0.80 (memory) and $0.70$ (network), while RCD achieves $AC@1$ of 0.50 (CPU). 
Notably, these methods also maintain high $AC@3$ ($\geq 0.80$) and $Avg@5$ ($\geq 0.69$) across most fault types. 
Their ability to place the true root cause at or near the top positions demonstrates practical utility beyond what $Avg@5$ alone conveys in small candidate sets.
However, methods like LiNGAM-PR/RW and MicroCause consistently underperform, often misranking non-faulty Pods such as the operator or gateway, with $AC@1$ near zero and $Avg@5$ below $0.25$, indicating limited applicability in LLM inference, mirroring their performance on microservice benchmarks.

\paragraph{Trace-based Methods}
In line with the utilized RCA benchmark and related work, trace-based methods are supplied tracing information about individual requests, including Trace ID, Span ID, Parent Span ID, service name, method name, operation name, status code, duration, and other detailed timing-related information.

Trace-based methods show limited effectiveness. 
TraceRCA yields $Avg@5$ around baseline ($0.28$–$0.55$) and $AC@1$ below $0.05$, suggesting it seldom identifies the true root cause as the top candidate. 
MicroRank remains at baseline across all fault types, with $Avg@5=0.20$ and zero top-1 accuracy, confirming that trace-only methods struggle in LLM inference contexts where the propagation of requests is not only minimal compared to multi-tier microservices, but also their handling is fundamentally different when recalling the request batching strategies of LLM inference systems.
Upon closer inspection, we noticed that the respective root cause has been regularly miss-attributed to the NGINX gateway, as this is one component where symptoms surface (e.g., due to request queue or latency buildup).

\paragraph{Multi-Source Methods}
In line with the utilized RCA benchmark and related work, multi-source methods are not only provided metrics and traces (as previously described), but also log data (composed of time, service name, and log message).
In addition, high-level statistics are computed from traces (error / success counts per service and operation over time windows) and logs (log counts per service over time windows) and, if supported, supplied to the respective method. 

Multi-source RCA approaches tend to outperform many metric-only and trace-only methods.
MM-BARO achieves $AC@1$ up to $0.77$ (memory) and $Avg@5$ of $0.87$ (network).
PDiagnose yields similar high $AC@3$ and $Avg@5$ across fault types, though its $AC@1$ remains lower (essentially zero across all fault types).
These results align with previous studies showing that integrating metrics and traces improves RCA performance. 
However, given the small candidate set, these near-ceiling $Avg@5$ scores are less conclusive than their substantial $AC@1$ and $AC@3$ improvements, which indicate more reliable top-ranked predictions.

\subsection{Key Findings on Misattribution Patterns}
\label{sec:experiment_results_misattribution}

\begin{figure}
    \centering
    \includegraphics{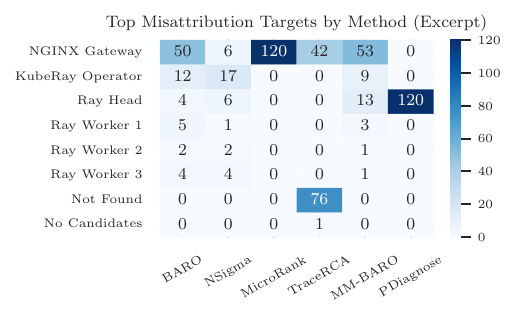}
    \caption{
    Misattribution counts of selected RCA methods across categories.
    Most approaches systematically assign faults to upstream components (NGINX Gateway or Ray head Pod) rather than the actual faulty Ray worker Pods.
    }
    \Description{
    Misattribution counts of selected RCA methods across categories.
    Most approaches systematically assign faults to upstream components (NGINX Gateway or Ray head Pod) rather than the actual faulty Ray worker Pods.
    }
    \label{fig:rca_misattribution_analysis}
\end{figure}

To deepen the understanding of the aggregate results shown in~\autoref{tbl:experiment_results}, we analyzed misattribution patterns of selected RCA methods (two methods for each category; for metric-based approaches we selected those with promising performance).
The corresponding outcomes are presented in~\autoref{fig:rca_misattribution_analysis}.
Across the evaluated methods, systematic errors in localizing anomalies became apparent.
In particular, five of the six approaches, namely BARO, NSgima, MicroRank, TraceRCA, and MM-BARO, regularly assigned responsibility to the NGINX Gateway, even when the injected faults originated from Ray workers.
This behavior points to a strong bias toward gateway-level components.
The remaining method, PDiagnose, avoided gateway misattribution but instead attributed failures to the Ray head Pod, indicating a similar tendency to blame higher-level components rather than the actual faulty Ray workers, since the Ray head Pod also takes part in load distribution of incoming requests.
Notably, for TraceRCA, the true root cause is frequently not even among the root cause candidates due to a filtering step applied internally within the method, leading to ``Not Found" situations.

Overall, none of the examined approaches identified the Ray worker Pods as the source of anomalies reliably ($AC@1$) across all fault types.
While these findings provide a clear picture of method-specific limitations under controlled experimental conditions, they also raise important questions for practical deployment.
In particular, the observed biases suggest potential inefficiencies in operational troubleshooting and highlight the need to reconsider how RCA techniques should be adapted to distributed microservice environments that increasingly include LLM-based inference services.
We turn to these implications in the following discussion.

\subsection{Discussion}
\label{sec:experiment_results_discussion}

\begin{figure}
    \centering
    \includegraphics{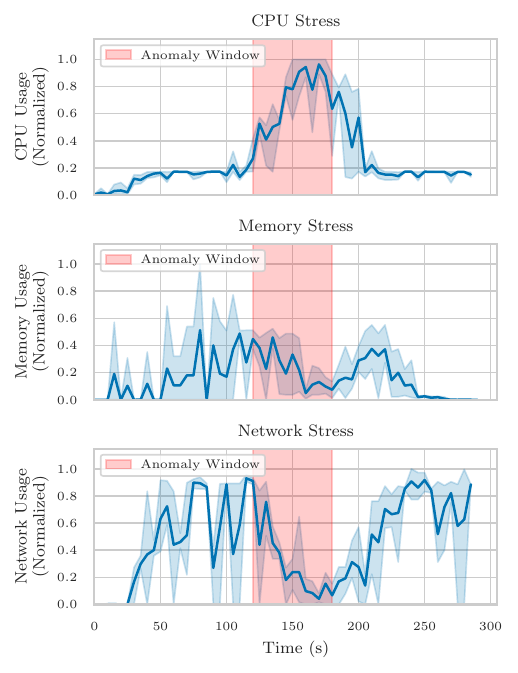}
    \caption{
    Effects of fault injections on system metrics.
    CPU hogs naturally increase the CPU usage over the time of injection.
    Network anomalies focusing on latency reduce overall communication for the time being.
    For memory, we first observe a decrease in usage, likely due to OOM conditions, before memory usage increases again when the Ray actor recovers.
    GPU throttling (not shown) led to no substantial changes in GPU utilization due to static allocations.
    }
    \Description{
    Effects of fault injections on system metrics.
    CPU hogs naturally increase the CPU usage over the time of injection.
    Network anomalies focusing on latency reduce overall communication for the time being.
    For memory, we first observe a decrease in usage, likely due to OOM conditions, before memory usage increases again when the Ray actor recovers.
    GPU throttling (not shown) led to no substantial changes in GPU utilization due to static allocations.
    }
    \label{fig:anomaly_injection_effects}
\end{figure}

The experimental evaluation revealed both the potential and the limitations of automated RCA in the context of LLM inference workloads.
In particular, the behavior of the system under different failure modes (cf.~\autoref{fig:anomaly_injection_effects}) offers insights into the observability requirements and the nature of fault propagation in such deployment scenarios.

CPU anomalies, such as synthetic CPU hogs, produced consistent and measurable effects on inference performance.
These anomalies degrade various latency-critical operations including request scheduling, tokenization, and data transfer between CPU and GPU.
As such, they manifest clearly in metrics such as response latency and token generation time, making them relatively easy to localize.
Furthermore, CPU anomalies can be injected in a reproducible and non-disruptive manner, facilitating their use in testing pipelines.
Memory-related anomalies, in contrast, often resulted in the abrupt termination of Ray actors due to out-of-memory (OOM) conditions.
Such failures caused specific model replicas to become temporarily unavailable until the actors were restarted.
This led to partial degradation of the deployment and, in some cases, to transient unavailability of the overall service.
Importantly, diagnosing such issues is more involved, as memory management and actor supervision are dependent on framework-specific mechanisms that are not always transparent or easily observable through standard system metrics.
This underscores the importance of including runtime-level logs and LLM-specific telemetry in the observability stack.
Network anomalies presented the greatest challenges in both fault injection and diagnosis.
In practice, we observed that some RCA methods often failed to converge during such failure scenarios, possibly due to incomplete telemetry, as Pod-level network chaos also affects sidecar data collection containers.
Manual intervention was frequently required to restart and recover from these experiments, limiting automation.
GPU anomalies, induced via power throttling, led to slowdowns in token generation and response times, reflecting degraded model execution.
Yet, RCA methods failed to detect them, as standard GPU metrics like utilization and memory remained unchanged due to static allocations, exposing a gap in current telemetry and RCA methodology.
Lastly, injecting HTTP-level delays at the API, as part of a series of exploratory additional experiments, caused unexpected Ray actor failures and unstable recovery behavior, which led to the decision to exclude these results from the paper.
Still, these early results highlight the fragility of distributed LLM systems when subject to perturbations in API-level behavior~\cite{DBLP:journals/corr/abs-2506-12320}, further complicating RCA.

Looking ahead, our findings underline the need for LLM-specific fault diagnosis.
Modern RCA methods like NSigma, CIRCA, BARO, and RCD perform well across most injected faults, while classical causal and trace-based approaches often misidentify non-faulty components.
This reflects the limits of traditional RCA in LLM contexts, where failures are subtle and tied to accelerator-bound execution as seen in the GPU throttling case.
Standard metrics alone are insufficient; signals like TTFT, token rates, and actor-level delays are essential.
Existing RCA tools designed for microservices represent a strong fundament to build upon, yet they require adaptation to handle the architectural and operational complexity of LLM-serving systems, especially under GPU-related failures like driver issues or device plugin faults.

\emph{Threats to Validity.}
Some factors may limit the generalizability of our findings.
We focused on common faults, such as CPU hogs, memory leaks, and network latency, and added a GPU throttling fault, yet excluding bugs or misconfigurations.
Fault parameters were adjusted to fit the higher capacity of our LLM setup, which may affect detection difficulty.
We slightly modified the RCA benchmark to support LLM telemetry and additional methods, potentially impacting comparability to the original evaluation~\cite{DBLP:conf/www/Pham00S025}.
Load generation followed a fixed LLM usage pattern, but may not capture the full spectrum of real-world deployments.
We deliberately based our evaluation of RCA methods on fundamental system telemetry (CPU, memory, network, GPU) rather than specialized application metrics that do not generalize, thereby aligning with related work, minimizing measurement and computation overhead, and avoiding bias introduced by an opinionated metric selection.
While these factors affect validity, the evaluation still offers valuable insights into the applicability of RCA for LLM systems.

\subsection{Gap Analysis and Observability Guidelines}
\label{sec:experiment_results_gap_analysis}

Our experiments show that many faults do not appear through CPU, memory, network, or GPU metrics alone, but surface via side effects in framework-specific behaviors such as Ray actor restarts or failed model loading.
These signals are often missed by standard observability stacks or hidden by default.
For example, Ray’s occasional use of IP-based direct Pod communication complicates network tracing and bypasses service meshes, hence requiring special configuration\footnote{\url{https://docs.ray.io/en/latest/cluster/kubernetes/k8s-ecosystem/istio.html}\LastTimeAccessed}.
We also had to adjust RayCluster placement strategies to reduce internal loopback and shared memory communication, which are difficult to observe out-of-the-box.

To close this visibility gap, observability for LLM systems must span multiple layers: system (CPU, memory, GPU), orchestration (Pod and container states), and application-level semantics (e.g., actor health, token timings).
This requires custom instrumentation in Prometheus, access to LLM-specific metrics like TTFT and E2E latency, and structured logging of process events.
Enriching telemetry with context, such as model names, actor IDs, and deployment topology, is essential for effective correlation.
While tracing can expose request flows, direct instrumentation is often impractical due to rapid framework changes and performance constraints.
Here, eBPF-based tracing (e.g., with DeepFlow) offers a non-intrusive alternative, capturing communication patterns and execution delays without modifying application code.

\begin{figure}
    \centering
    \includegraphics{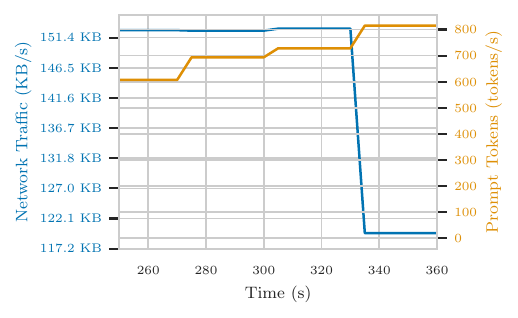}
    \caption{
    Contrasting application-specific signals after a memory anomaly: network traffic decreases while prompt tokens per second increase, illustrating counterintuitive effects that are difficult to interpret without deep knowledge.
    }
    \Description{
    Contrasting application-specific signals after a memory anomaly: network traffic decreases while prompt tokens per second increase, illustrating counterintuitive effects that are difficult to interpret without deep knowledge.
    }
    \label{fig:network_traffic_vs_prompt_tokens}
\end{figure}

A modern observability strategy must hence balance low overhead, deep insight, and adaptability.
This calls for combining cloud-native observability with LLM- and framework-specific knowledge to meet the demands of LLM inference.
However, even with access to deep insights, identifying which telemetry signals are relevant is rarely straightforward and must be weighed against practical considerations such as the frequency of framework changes (in case of application-specific telemetry signals) and the overhead of initial signal selection and instrumentation.
Even when such signals are available, interpreting their interrelations is not always trivial, since correlations may be complex or misleading.
To illustrate this challenge, we contrast two application-specific metrics in~\autoref{fig:network_traffic_vs_prompt_tokens}, highlighting how their relationship is far from obvious: Following an injected memory anomaly that triggered an OOM failure in a Ray worker Pod, we observe a reduction in reported in-cluster network traffic.
Concurrently, the overall throughput in terms of prompt tokens per second increases, since the workload is redistributed to the remaining worker Pods, resulting in more batched request processing and prolonged individual request latency, thereby elevating this metric.
Without detailed knowledge of the framework’s internal behavior, however, such counterintuitive relationships are difficult to interpret, both for human operators and for RCA methods.

\subsection{Toward RCA-Aware AI Infrastructure}
\label{sec:experiment_results_toward_rca_aware}

To support timely and reliable RCA in LLM inference systems, observability must evolve into infrastructure-aware analysis.
Traditional RCA assumes linear causality and stable component behavior, but LLM workloads rely on GPU-bound actors, shared memory, distributed object stores, and cooperative scheduling.
RCA methods must be adapted to incorporate these architectural patterns.
For instance, a failing Ray actor may lead to retries, rescheduling, or queuing delays; all symptoms of an upstream fault.
Without fine-grained visibility into actor lifecycles and GPU behavior, RCA methods risk misattribution.
Effective RCA should integrate awareness of LLM-specific abstractions such as actor execution, KV cache usage, and model stages.
It must also incorporate GPU telemetry, model-serving metrics, and client-facing API data.
Handling degraded states and partial recovery patterns is essential.
Logs, metrics, and traces must be correlated across system layers and under time constraints.
Generally, rather than relying solely on generic signals, RCA systems should embed declarative knowledge of LLM behavior: a precise definition of a healthy actor, a latency decomposition, and faults propagation details.
This enables not only accurate fault localization but also actionable remediation, which may include the automated execution of recovery actions.

\begin{figure}
    \centering
    \includegraphics{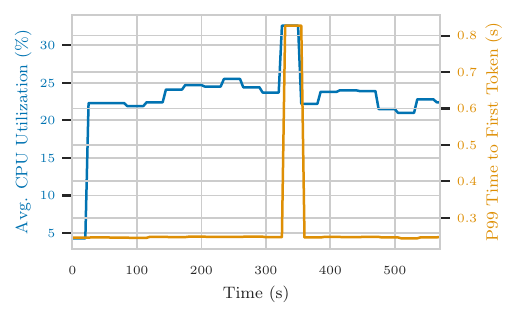}
    \caption{
    Self-stabilization experiment: injection of a CPU Hog fault causes increased CPU utilization and TTFT spikes; RCA identifies the faulty worker Pod and triggers a restart, restoring stable performance while also highlighting challenges in interpreting TTFT as a capacity indicator.
    }
    \Description{
    Self-stabilization experiment: injection of a CPU Hog fault causes increased CPU utilization and TTFT spikes; RCA identifies the faulty worker Pod and triggers a restart, restoring stable performance while also highlighting challenges in interpreting TTFT as a capacity indicator.
    }
    \label{fig:cpu_util_vs_time_to_first_token}
\end{figure}

As an illustrative case, we highlight the self-stabilization experiment shown in~\autoref{fig:cpu_util_vs_time_to_first_token}, conducted as part of an exploratory series of experiments.
When a CPU anomaly was introduced into a Ray worker Pod, overall CPU utilization rose sharply and the Time-to-First-Token (TTFT) metric spiked, indicating degraded inference performance.
The anomaly detection component, implemented through lightweight alerting rules, in combination with a selected RCA method from our earlier analysis, correctly identified the affected Pod as the likely root cause and automatically initiated a restart.
During the restart phase, the Pod was removed from Kubernetes’ request routing, which allowed CPU load and TTFT values across the remaining healthy Pods to stabilize.
Once the restart was completed, the recovered Pod resumed normal operation.
This example illustrates how observability, coupled with RCA, can enable automated corrective action.
At the same time, it highlights challenges in signal interpretability: although the system’s effective capacity was temporarily reduced during recovery, this effect was not reflected in TTFT, likely due to how vLLM defines and implements the metric—namely, by measuring only for requests already in execution rather than those queued for processing.

Generally, RCA-aware observability must align closely with system semantics.
This points toward a broader shift: from anomaly detection to context-driven inference, grounded in the operational logic of AI workloads.
Such alignment is key to building reliable and maintainable AI infrastructure.

%% file: sections/6_related_work.tex
\section{Related Work} 
\label{sec:related_work}

We revisit different areas of related work in the following.

\subsection{Traditional RCA in Distributed and Cloud-Native Systems}

Classical RCA research spans model-based, statistical, and graph-centric segments.
Model-based systems encode expert rules or dependency maps to trace symptom propagation.
Their accuracy depends on manual upkeep and quickly degrades as microservice topologies evolve.
Statistical approaches treat every metric as a time series, ranking potential culprits by correlation or causality scores.
Seer~\cite{DBLP:conf/asplos/GanZHCHPD19} and LOUD~\cite{DBLP:conf/icst/MarianiMPRX18} exemplify this line: both flag resources whose usage trends precede service-level-objective violations.
These methods scale to thousands of metrics but often mistake workload shifts for faults.
Graph-based RCA methods like MonitorRank~\cite{DBLP:conf/sigmetrics/KimSS13}, Microscope~\cite{DBLP:conf/icsoc/LinCZ18}, and MicroRCA~\cite{DBLP:conf/noms/WuTEK20} add structural context. 
They build service or metric graphs and apply random-walk or centrality algorithms to pinpoint nodes most consistent with observed faults.
Graph reasoning captures multi-hop fault propagation, relying on complete traces. 
Occasional false-detection happens when the graph is sparse or changing.
Collectively, these works laid the foundations for microservice RCA, usually assuming conventional web service workloads.
More recent approaches~\cite{DBLP:conf/nips/IkramCMSBK22,DBLP:journals/corr/abs-2406-05375} extend classical model-, statistical-, and graph-based methods with new AI-driven techniques, while still facing similar challenges (e.g. scalability and evolving system topology).

\subsection{LLMs as Tools for RCA}

In recent times, LLMs have also been used to automate parts of incident diagnosis.
Microsoft’s study on using GPT models for RCA in cloud operations demonstrated that LLMs can generate plausible cause hypotheses and mitigation suggestions directly from incident data~\cite{DBLP:conf/icse/AhmedGBZZR23,DBLP:conf/sigsoft/ZhangGBWM0R24}.
More structured agent frameworks such as RCAgent~\cite{DBLP:conf/cikm/WangLZZWYF0W24} combine LLM reasoning with external data access tools, enabling more accurate and autonomous root cause identification.
Some recent efforts have begun exploring the integration of LLMs themselves into AIOps pipelines~\cite{DBLP:conf/IEEEscc/LimW24}, for example, using GPT-based models for log summarization, incident explanation, or query routing~\cite{DBLP:conf/kdd/ZhongMLLLZWLW24}.
These approaches show potential in reducing operator effort and providing semantically rich insights, but face challenges such as prompt sensitivity, non-determinism, and difficulty integrating LLM outputs into deterministic workflows~\cite{DBLP:journals/corr/abs-2501-12461}.
Other recent efforts have explored LLM-based agents and benchmarks for RCA.
Flow-of-Action~\cite{DBLP:conf/www/PeiWLLLHKZCLXP25} combines an LLM-driven multi-agent framework with standard operating procedures to guide diagnosis and reduce hallucinations. 
This approach nearly doubled fault-localization accuracy (from $\sim$35\% to 64\%) compared to a vanilla ReAct agent. 
Complementing such methods, OpenRCA~\cite{DBLP:conf/iclr/XuZZHZLPHZ025} introduced a public benchmark of 335 real failure cases (68GB of logs, metrics, traces) to evaluate LLMs’ diagnostic abilities.
Results show current models can only handle the simplest cases: even with a specialized agent, the best model solved $\sim$11\% of failures, highlighting significant room for improvement in LLM-powered RCA.

\subsection{Toward Failure Handling in LLM Inference Services}

LLM inference forms a distinct infrastructure class.
Engines like DeepSpeed, vLLM, and Llama.cpp are optimized for memory and throughput, using mechanisms like continuous batching and paged attention.
These optimizations introduce novel failure modes linked to performance tuning and cross-platform complexity.
While prior work has studied training-stage bugs~\cite{DBLP:conf/hpca/KokolisKHKMMDSS25,DBLP:journals/corr/abs-2502-05413}, only recently has inference become a focus.
Liu et al.~\cite{liu2025lookbugsllminference} provide the first large-scale bug analysis in LLM inference, identifying 929 bugs across five engines and categorizing 28 root cause types.
Notably, over 35\% of issues present as non-crash symptoms, such as incorrect outputs or hangs, making them harder to detect and diagnose.
Another recent analysis~\cite{DBLP:journals/corr/abs-2506-12320} of HuggingFace Transformers and vLLM identified API misuse as the leading root cause ($\sim$40\% of bugs), marking a shift from algorithmic bugs in earlier Deep Learning frameworks to interface-level problems. 
It also found that the majority of bugs escape existing tests (often due to inadequate test cases).
This shift has spurred interest in specialized monitoring and testing tools~\cite{DBLP:conf/iwqos/XuXC25}.
Failures like mis-encoded tokens or OOM errors motivate custom health checks and logs in LLM-serving systems.
Overall, ensuring LLM inference reliability requires RCA methods informed by engine internals and failure modes.

%% file: sections/7_conclusion.tex
\section{Conclusion}
\label{sec:conclusion}
As large language models are increasingly integrated into real-world applications, ensuring the reliability of their serving infrastructure becomes essential.
This paper investigated whether existing methods for identifying the root cause of system failures, originally developed for microservice-based architectures, continue to be effective when applied to modern language model inference systems.
Using controlled fault injection on an LLM inference stack aligned with best practices, we found that existing techniques can achieve promising results while often revealing gaps related to specifics of LLM inference stacks.
These failures arise due to core architectural differences.
LLM serving systems rely on specialized hardware, dynamic batching, and shared memory, breaking assumptions like statelessness, traceable communication, and uniform telemetry.
As a result, traditional RCA tools occasionally misattribute anomalies or miss faults entirely.
We identified key observability requirements for LLM inference, including hardware-aware metrics, internal queuing and scheduling visibility, and representation of asynchronous execution, which may enable actionable remediation.

Our findings suggest that diagnosing failures in generative model infrastructure requires revised analysis methods and expanded observability coverage.
This work provides an early empirical foundation for adapting failure diagnosis tools to meet the demands of increasingly complex, performance-sensitive machine learning systems.
In the future, we plan to expand our experiments to larger models and clusters, as well as to diverse model parallelism setups.